\documentclass[epj,referee]{svjour}
\usepackage[dvips]{graphicx}

\def\jpc#1#2#3{J.~Phys.~Chem.~A~{\bf #1},\ #2\ (#3)}
\def\jcp#1#2#3{J.~Chem.~Phys.~{\bf #1},\ #2\ (#3)}
\def\cpl#1#2#3{Chem.~Phys.~Lett.~{\bf #1},\ #2\ (#3)}

\def\pra#1#2#3{Phys.~Rev.~A~{\bf #1},\ #2\ (#3)}
\def\prl#1#2#3{Phys.~Rev.~Lett.~{\bf #1},\ #2\ (#3)}

\def\jpb#1#2#3{J. Phys. B: At. Mol. Opt. Phys. {\bf #1},\ #2\ (#3)}

\newcommand{\beq}{\begin{equation}}
\newcommand{\eeq}{\end{equation}}

\begin{document}
\title{Editorial: Quo vadis, cold molecules?}

\author{John Doyle$^1$,  Bretislav Friedrich$^{1,2}$, Roman Krems$^3$ and Fran\c{c}oise Masnou-Seeuws$^4$ %
}                     
%
\institute{$^1$ Department of Physics, Harvard University, Cambridge, MA 02138, USA
\\$^2$ Fritz Haber Institute of the Max Planck Society, D-14135 Berlin, Germany
\\$^3$ ITAMP, Harvard-Smithsonian Center for Astrophysics, Cambridge, MA 02138, USA
\\$^4$ Laboratoire Aim\'e Cotton, CNRS, B\^at. 505 Campus d'Orsay,
91405 Orsay Cedex, France}
\date{Received: date / Revised version: date}

%
\abstract{ The present issue gives a snapshot of the rapidly developing field of ultracold polar molecules.}
\PACS{
      {33.80.Ps}{Cold atoms and molecules} 
     }

\authorrunning{Doyle, Friedrich, Krems, and Masnou}
\titlerunning{Quo vadis?}
\maketitle

\section{Introduction}
The idea to produce this Topical Issue came to us during a meeting in
Cambridge last January, the ITAMP-CUA Workshop on Ultracold Polar 
Molecules\footnote{ITAMP is the Institute for Theoretical, Atomic, Molecular 
and Optical Physics;
CUA is the Harvard/MIT Center for Ultracold Atoms.}.
The ultracold polar weather that beset Cambridge at that time contributed to
a congenial and, indeed, warm atmosphere in the halls and hallways of the
workshop's venue, the Harvard-Smithsonian Center for Astrophysics.

\textbf{Place picture}

The present collection of papers is a snapshot of a very rapidly developing
field. There are now at least nine different techniques used to produce cold
molecules, a number that increased from zero over the past eight years.

As \emph{cold}, we designate molecules with translational temperatures
between $1$ and $1000$ mK. Such molecules can be easily manipulated with
electromagnetic fields and trapped. As \emph{ultracold}, we designate
molecules whose translational temperatures are typically less than $1$ mK;
such molecules are distinguished by a single partial-wave scattering.

\subsection{Why cold polar molecules?}
Currently, the goal of many researchers is to produce dense samples of
ultracold polar molecules. This goal is shared by colleagues with diverse
backgrounds, mainly in physical chemistry, 
chemical physics and in atomic and molecular physics. Why such an
intense interdisciplinary interest? Molecules are different from atoms -
certainly more complicated - and one might say ``more interesting:'' they
offer intriguing properties which are not available with atoms, such as
body-fixed electric dipole moments. Hence the emphasis on \emph{polar}.

As atomic physics has moved beyond the trapping and study of dilute gases of
simple atoms, a new many-body physics has emerged. Similarly, fundamental 
physical
chemistry is moving in new directions, 
becoming increasingly preoccupied with controlling chemical
reactions and collisions in general. In both research areas dense samples of
cold molecules are called for - and wanting. Once these become available,
much is expected to happen: there is the promise of the discovery of new
phases of matter, ultracold chemistry, and even a robust quantum computer
made out of molecules suspended in free space. These far-reaching research
goals and the thrill of the challenge drew a group of about seventy
researchers through the ultracold weather to talk about polar molecules.
This volume is not a summary of that meeting (please see the Workshop's web
page for that, http://itamp.harvard.edu/polar/polar.html) but a series of
peer-reviewed articles fostered by that meeting. We hope that you will find
this collections of papers useful and inspiring.\vspace{0.03in}

\subsection{How to create cold polar molecules?}

The discovery of methods for cooling, trapping, and manipulating atoms has
generated a revolution in atomic physics. In the past few years many
researchers have chosen to pursue the creation and study of ultracold polar
molecular gases. Cooling  molecules is more difficult than cooling
atoms. The complex molecular internal level structure precludes a simple
extension of laser cooling to molecules (although work is continuing in this
area and the proposals to laser-cool metal hydrides look promising 
\cite{dirosa,bsg96}). Despite hurdles, considerable progress has been made and
we are in a period of very rapid world-wide growth in this new field. In
1998, the first trapping of neutral polar molecules, CaH, in
a magnetic field was reported \cite{cahnature}. Since then, several groups have made
advances in the production of cold polar molecules by a variety of methods.
The creation of homonuclear (bosonic) alkali dimers via photoassociation or
Feshbach resonances is now being followed up with application of the same
techniques to produce polar bi-alkali dimers; recently Feshbach resonances
in dual-species traps have been found \cite{szs04,igo04}. Ground-state
(or at least triplet ground state)
bi-alkalis have been produced by photoassociation \cite{gould,kss04,mtc04,hkb04}. 
The methods of photoassociation and Feshbach-resonance linking can be termed
``indirect'' as the molecules are formed from pre-cooled atoms (see Section
2 of this Editorial). There are more than twenty 
groups pursuing the indirect methods
and some are turning their attention from homonuclear to polar molecules. 
The advantages and drawbacks of the indirect methods are described
in Section 2. 

 ``Direct'' methods are based on 
cooling preexisting molecules. By current count there are at least fifteen groups
pursuing either buffer-gas cooling 
\cite{cahnature,csv04,peters03,demilleprivate,dbn04}, electric-field pulsed beam slowing 
\cite{bm03,bhl04,ed_hinds,tbh04,hgould}, light-field slowing 
\cite{bbf04}, laser cooling \cite{dirosa}, 
counter-rotating-nozzle slowing \cite{gh99}, single collision scattering
\cite{evc03}, or beam skimming with a guide \cite{rjr03,nda03}. 
All of these direct methods start with relatively hot molecules, typically from a molecular
beam source, and employ some combination of slowing, cooling and trapping as
an initial step toward creation of ultracold polar molecular gases (see
Section 3 of this Editorial). The advantages of (many of) the direct
approaches include wide applicability  and large yield (CaH, OH, and NH$_{3}$ have been
already trapped and a wide variety of molecules, including benzene, have been
slowed).
 A disadvantage is that many of these methods generally do not
immediately yield the very low translational temperatures, typically in the
microkelvin range, which are attainable with the indirect methods. It will be
necessary to bridge the ``temperature gap'' between the initial loading
temperatures of the direct methods (1-1000 mK) and the desirable ultracold
regime $<1$~mK. There are a wide variety of proposals on how to do this,
including direct evaporative cooling, sympathetic cooling with laser- or
evaporatively cooled atoms, or direct laser cooling. The application of
these techniques will generally require starting with 
a large number of molecules. It will
also require learning a great deal of new physics. 

The payoff should be worth the effort. 
Most prominently, the
large electric dipole moments of polar molecules produce a strong
interparticle interaction that can be exploited for quantum computing or
as a source for direct BCS pairing 
\cite{demille02,ssz00,bdg02,csc04,bdl04,kleppPT04}. 
Also, precision spectroscopy of heavy dipolar molecules in
searches for an elementary-particle electric dipole moment (EDM) \cite{ed_hinds,tbh04} 
holds tremendous promise for reaching physics beyond the Standard Model.
Producing larger numbers of colder molecules will naturally lead to
improvements in these efforts. Beyond these anticipated phenomena, there
awaits a \emph{terra incognita}. What are the possibilities for chemistry at
low temperature? Will there be applications to clocks? Are there new
collisional phenomena? These are some of the questions that are on the
table. Just as we are now seeing phenomena in the ultracold atom field not
dreamt about twenty years ago when BEC\ research was launched, we expect
that new and exciting molecular phenomena we can't now imagine will emerge
ten to twenty years hence.

\section{Making ultracold polar molecules from cold atoms}
\subsection{Photoassociation and radiative stabilization}

Laser cooling techniques cannot be easily applied to molecules \cite{bsg96,djeu81}, 
which are complex multi-level systems. One can circumvent
 this difficulty by first cooling  the constituent atoms, then making excited state molecules 
by photoassociation of  atom pairs with laser light, following the scheme proposed 
in 1987 by Thorsheim {\it et al.}~\cite{thorsheim87} and discussed 
in multiple reviews  \cite{weiner99,bahns00,masnou01,bill_review}. 
However,  molecules thus obtained are short-lived and decay 
by spontaneous emission, 
most often dissociating back into atoms. In order to produce 
stable molecules,   
population transfer from a bound level $v'$ of the excited electronic state to 
bound levels $v$ of the ground (or lower triplet) electronic 
state must occur. This stabilization process  involves either  spontaneous emission, or induced emission with a second laser. 
The advantage of the photassociation 
method is that 
one makes ultracold molecules at the same translational temperatures
as the precursor atoms (below 1 mK).
A drawback
is that the  molecules are often 
formed in excited vibrational levels of the electronic ground state, so that they are not vibrationally cold.
 Also,  in many cases,  the stable molecules are formed in the lower 
triplet state rather than the ground state
and they can be captured due to their low temperature
either in a magnetic trap  \cite{bahns00,masnou01}
or in a dipole trap at the focus of a CO$_2$ laser beam \cite{knize}. 
 
The first stable ultracold molecules that were formed, detected by photoionization and 
time-of-flight selection of the molecular ions,   
 were homonuclear alkali dimers 
\cite{fioretti98,takekoshi98,nikolov99,nikolov00,gabbanini00,dion01,fatemi2002}. 
Several mechanisms were identified as the stabilization step, all 
associated with specific properties of the potential curves in the excited electronic state.
The progress and development of 
new schemes appear to be strongly dependent on accurate spectroscopic information about 
the molecules and on vigorous symbiosis between theory and experiment (see the reviews \cite{bahns00,masnou01} and references therein).

As for the formation of heteronuclear dimers, 
there is an added complication of using a double MOT. 
This issue is ``merely" technical and has been solved by 
many groups. Regarding the photoassociation itself,  
it was predicted
\cite{wang98} that the photoassociation step would be less efficient for
heteronuclear dimers
than for homonuclear molecules. 
The heteronuclear molecular potentials in the excited electronic state vary asymptotically as  $R^{-6}$, 
while for homonuclear molecules the  potentials correlating with the first excited 
asymptotic limit have a long range  
$R^{-3}$ behaviour due to  the resonant exchange interaction. Since the photoassociation of ultracold atoms occurs  at 
large distances, the $R^{-3}$ potentials, which support bound levels with larger extension, are more favourable. However, 
the stabilization step is quite
efficient for hetereonuclear dimers, due to the similar $R^{-6}$ asymptotic 
behaviour of both the  ground and excited potential
curves. This makes bound-bound transitions more probable in heteronuclear molecules 
than in homonuclear dimers. 
KRb, RbCs and KCs were predicted to be promising candidates
\cite{wang98}. After successful detection 
of a signature of stable molecules, molecular ion production
\cite{bigelow99,grimm99}, it took six years to make  
ultracold heteronuclear dimers with clear identification of the production mechanisms.
The following molecules were first formed and state-identified in 2004: RbCs by the Yale
group~\cite{kerman04a,kerman04b} and KRb by the UConn\footnote{UConn is 
the University of Connecticut.}
group~\cite{gould} and by the Sao  Paulo group \cite{mancini2004}.
Contributions from the first two  groups are presented in this issue,
which contains
more than ten papers on the subject of photoassociation. 

The UConn  group's paper by {\bf Wang {\it et al.}} (pp xx-yy) presents
a detailed photoassociation spectroscopy of KRb, formation of molecules in the
ground triplet state a$^3\Sigma^+$, and detection of stable molecules using a two-photon ionization scheme. A wealth of
information is thus available on the molecular levels in the range of 95
cm$^{-1}$ below the K(4s) and  Rb(5p$_{1/2}$) asymptotic limit, 
elucidating the interactions and perturbations between various vibrational series.
Such a resonant structure was known to be important for the
efficiency of the stabilization step (resonant coupling mechanism \cite{dion01}). The
paper shows that it can also be a loss path, through predissociation,
since  a photoassociated molecule could rapidly  dissociate before
emitting a photon that would stabilize it. Information  on potential
curves and transition moments can be obtained from the spectra, yielding
useful data to design new formation schemes. The stable KRb molecules can be
trapped in a magnetic field, as expected for triplet state molecules, which was
already demonstrated for Cs$_2$ \cite{cs2} and Rb$_2$ \cite{rb2}. 

The paper of {\bf Bergeman, Kerman,
Sage, Sainis,} and {\bf DeMille} (pp xx-yy) investigates
the possibility  of transferring RbCs stable molecules from the lower
triplet state a$^3\Sigma^+$ to the ground state X$^1\Sigma^+$. 
Stable molecules are created as in previous experiments 
\cite{kerman04a,kerman04b}. We should note the large value of the 
densities (close to 10$^{12}$ cm$^{-3}$) for both species, the
temperature of 75 $\mu$K, and the very large number of Cs atoms
attained in these experiments.
Stable molecules are formed in a bound   
level of the lower triplet state and they are re-excited  by a photon
into a bound level of an excited electronic state
correlating with the Rb
(5s) + Cs (6p) asymptotic limit. From an analysis of the bound-bound spectrum, data
are extracted on the excited potential curves and on the coupling between
various channels. Using this information, the best schemes to transfer 
population from a bound level in an excited electronic state (or in a
mixture of such excited  states, due to coupling) to a bound level of the
ground X$^1\Sigma^+$ state are discussed. From the Franck-Condon
factors, it is shown that populating the $v=0$ and $v=1$ levels 
is a realistic goal, and that making translationally and rotationally cold
molecules that are also in the ground vibronic state 
should be feasible.

The same problem is addressed theoretically in the
paper by {\bf Kotochigova, Tiesinga,} and {\bf Julienne}  (pp xx-yy)  for the KRb   
molecule. The potential curves and transition dipole matrix elements
are determined by {\it ab initio} calculations. 
A wide range of transition moments is thus  
available. The starting point is a pair of K(4s)  and Rb(5s)  atoms,
spin-polarized and therefore interacting within the 
a$^3\Sigma^+$ state. The photoassociation laser creates a resonance  with a level $v'$ in the excited
state (2 0$^+$). The latter  is such that the transition dipole moment to a vibrational
level of the ground X $^1\Sigma^+$ is also large. This is due to the fine  
structure coupling in the excited state, where the singlet and triplet
states are mixed, so that radiative transitions to both ground and
lower triplet states are allowed. The authors note that this is not possible
for homonuclear molecules, where the $u-g$ symmetry would forbid  one of
the transitions. Thus, it is a big advantage of the heteronuclear systems.
The authors propose  a two-color Raman photoassociation experiment in which 
the first photon is slightly detuned from a photoassociation resonance, while the
second photon drives the transition from the (2 0$^+$) level to a
vibrational level of the ground state: the numerical calculations  show
that the latter can be rather low ($v<30$) and propose optimal paths. Using the
computed values for the permanent dipole moments in the ground state, the 
authors discuss the stability of heteronuclear molecules relative to 
thermal blackbody radiation, which would cause cascading from one
vibrational level to another in the same electronic state.
However, the lifetimes are found to be much longer than the typical duration 
of an experiment, in agreement with earlier 
calculations \cite{bill_rq1}.

Is it possible to go beyond the favorite molecules RbCs and KRb?  The   
theoretical paper by {\bf Azizi, Aymar,} and {\bf Dulieu} (pp xx-yy) 
assesses
photoassociation  and cold molecule formation rates, which
are computed for all the heteronuclear dimers composed of either Rb or Cs
atoms. As in the previous paper, the rates are computed from numerical
vibrational wavefunctions and the integral over the dipole
transition moment rather than estimates of the probability for 
vertical transitions. The results are compared to the homonuclear Cs$_2$ rates,  
using a rescaling factor for the differences in reduced mass and atomic
Rabi frequency. It is shown that the photoassociation rates are reduced by
a factor $\approx 30$  for a pair of identical atoms compared to a pair
of different atoms. However, since 
for heteronuclear dimer molecules the excited potential curves 
have the same $R^{-6}$ asymptotic behavior as the ground or lower triplet
curves, the radiative decay to bound levels is faster than in
homonuclear molecules, making  the stable molecule formation rates only one
order of magnitude smaller than for Cs$_2$. Another conclusion is that the
lighter species (LiRb, NaRb, LiCs, NaCs) should be much better candidates 
than previously thought, therefore encouraging experiments in that
direction.

As for NaCs, detailed information, necessary to design new schemes, is now
available from the extensive spectroscopic work of {\bf Docenko, Tamanis,
Ferber, Pashov, Kn\"ockel,} and {\bf Tiemann} (pp xx-yy), from Riga and Hannover. In the   
paper published here, the molecular spectroscopy experiment is described
and a very accurate determination of the X$^1\Sigma$ potential curve is
presented, by fitting an analytical function to the measured energies of 
more than 2000  rovibrational levels. The long range coefficients such as
$C_6$ can thus be determined very precisely. 
The spectrum provides information on the shallow 
a$^3\Sigma^+$ potential
and the authors
estimate the scattering lengths for both potentials.

Another candidate appears to be the LiH molecule, as discussed in 
the paper of {\bf Taylor-Juarros, C\^ot\'e,} and {\bf Kirby} (pp xx-yy). 
The dipole moment of the alkali hydride molecules can be larger than that 
of heteronuclear alkali dimers.
As in the two theoretical papers described above, the    
authors compute the photoassociation rates from a continuum level to a  
bound level $v'$  of the A$^1\Sigma^+$ excited state, followed by
stimulated emission rates to a bound level $v$  of the ground X$^1\Sigma^+$
state, looking for the best $v,v'$ combination for a stimulated Raman   
photoassociation experiment.  Important  
formation rates are found for stable molecules. An  advantage of the alkali hydrides is that 
due to the large dipole moment in the X$^1\Sigma^+$ state (typically one
order of magnitude larger than in the KRb molecule discussed above) the    
stable molecules formed in an excited $v$ level decay by spontaneous emission to
the ground vibrational level  $v=0$ with relatively short lifetimes.

The Raman scheme is also discussed by {\bf Stwalley} (pp xx-yy) with the aim of
converting  molecules formed by Feshbach resonances into  ground state   
molecules. The initial state is considered to be a loosely bound level
of $^3\Sigma^+$ symmetry rather than the continuum, corresponding to recent
experiments in condensates with magnetically tuned Feshbach resonances. It could
also be populated by photoassociation followed by spontaneous emission.
Besides KRb, the molecules discussed from a general perspective are  LiNa,
NaK, NaRb, and RbCs. The region of the spectrum where detailed studies
should be performed is indicated. 

Up to now, photoassociation has been used to populate bound levels in
the electronic states correlating with the first excited atomic limit. 
It was often suggested that highly excited molecules at intermediate and long range 
were also good candidates \cite{bill_rq2,almazor99,laue2003}. 
Alternatively, the initial state can also be a continuum  
state of an excited potential curve: a very interesting experiment  was
recently performed  by the ENS\footnote{ENS is Ecole Normale Superieure.} 
group in Paris \cite{leonard2003}. Starting
from spin-polarized metastable helium ($^4$He) atoms at very low
temperatures (2 to 30 $\mu$K), bound levels of electronic states correlating with
the $2^3$S$_1$ + $2^3$P$_0$ asymptotic limit are populated by
photoassociation. Details are presented here by {\bf Kim, Rapol, Moal,   
L\'eonard, Walhout,} and {\bf Leduc} (pp xx-yy). Such molecules should decay by Penning
ionization, but for some symmetries (0$_u^+$, 1$_u$)  the existence of  
long range wells constrains the relative motion to very large  
interatomic separation, with the inner turning point around 150 a$_0$ (while the
outer turning point can be as far as 1100 a$_0$).  Penning ionization
is therefore inhibited, as photoionization was inhibited in Ref. \cite{bill_rq2}. 
The novelty of the experiment is in the    
calorimetric detection of the photoassociation resonance: when the laser
frequency is tuned to a molecular transition, creating a bound
photoassociated molecule, the latter decays by spontaneous emission, yielding
a pair of atoms that heat the cloud of metastable atoms. This increase of
temperature can be detected by optical methods. Very accurate
photoassociation spectra are thus produced, with line widths of a few MHz.  
Excellent agreement is obtained between the observed spectrum and
calculations based on  long-range potentials. Future work will explore 
the magnitude of the scattering length by 
accurate photoassociation spectroscopy. 

All the papers described so far have treated the laser coupling in 
the photoassociation reaction in a perturbative way. In fact, laser light can
also be used to manipulate molecules, control the photoassociation
reaction or tune the scattering length. The concept of an 
optically induced 
Feshbach resonance  is very useful for understanding such physical
processes. We will discuss optically induced 
Feshbach resonances in this section and magnetically induced Feshbach resonances 
in Sec. 2.1.  

The theoretical paper by the Orsay group of {\bf Luc-Koenig, Vatasescu,} and
{\bf Masnou-Seeuws} (pp xx-yy) discusses the possibility of replacing cw lasers by
a series of chirped laser pulses which optimize  both the
photoassociation rate and the formation of stable molecules in low
vibrational levels. A time-dependent non-perturbative treatment of the
photoassociation reaction in a cold cesium sample has been proposed
earlier in 2004 \cite{luc2004}. In the present paper the authors discuss
the choice of the parameters of the chirped pulses (intensity, duration,
linear chirp rate). The variation of the laser frequency  allows one to sweep
through
several optical Feshbach resonances, determining a photoassociation window
within which  complete adiabatic population transfer from the ground
electronic state to bound levels of the excited state is taking place.  A
large number of vibrational levels can be populated coherently, so that
with a reasonable repetition rate of the laser pulses, the number  of
molecules formed per unit time is expected to be much larger than with a
cw laser of equivalent power. Besides, due to scaling laws on the
vibrational periods of long-range molecules, revival phenomena can occur
for the vibrational motion in the excited state. It is therefore possible to
compute the linear chirp parameter in order to create, within the excited state,
a wavepacket focused 
at the inner turning point of a long range well.
This wavepacket has a large Franck-Condon overlap with low vibrational
levels of the ground state, so that a second laser pulse could create
vibrationally cold stable molecules.
The method is similar to the Raman scheme discussed in many papers of this
issue but uses pulsed lasers. The advantage of short pulses is that
molecules in highly excited vibrational levels can be transferred to the
ground level in very short times, shorter than collisional relaxation or
characteristic dissociation times \cite{koch2004}.

Manipulating the scattering length with a photoassociation laser has 
been proposed in several 
theoretical papers \cite{fedichev96,bohn97,kokoouline2001}, and achieved in the experiments of Refs. \cite{fatemi2000,theis2004}.
The paper by {\bf Koelemeij} and {\bf Leduc} analyzes 
the possibility of using photoassociation to determine the scattering length of metastable helium (through accurate 
measurement 
of the binding energy of the last bound level of the 
ground state) and to manipulate it in order to control the dynamics of metastable helium condensates. 

  A generalized theory of photoassociation in quantum degenerate gases 
is presented in this issue by {\bf  
Mackie, Dannenberg, Piilo, Suominen,} and {\bf Javanainen} (pp xx-yy). 
In particular, the authors discuss photoassociation (together with magnetic Feshbach resonances, see below Section \ref{ssec:feshbach}) in mixed 
Fermi-Bose gases. They report a
numerical observation of large-amplitude atom-molecule
oscillations similar to those occurring in
pure Bose gases. 
The authors also discuss 
the prospects for   fermion atom - fermion molecule pairing
and explore the transition from
triatomic molecules to correlated atom-molecule
pairs.

We have already mentioned that photoassociation 
leads to 
long-range molecules. 
The vibrational motion of long-range molecules is determined by long-range interatomic forces which can be effectively 
described by theoretical models such as generalized quantum defect theory.
 The cold molecule research field is therefore linked to the field of Rydberg atoms, 
and many concepts are similar 
(we have mentioned earlier scaling laws and revival phenomena) 
with differences arising from the $R^{-n}, n=3,6$, rather than $R^{-1}$ asymptotic behaviour. 
This very fertile topic is represented here by the paper of {\bf Gao} (pp xx-yy), which reports systematics of rotational energy levels of diatomic molecules in the vicinity of
a dissociation threshold. A new classification scheme of long-range states 
of diatomic molecules is proposed. The study should be helpful for 
the analysis and identification of molecular spectra required for 
the photoassociation experiments, and for discussion of the determination of the scattering length from 
spectroscopy of the last vibrational levels 
(see the two papers on metastable helium quoted above, Ref. \cite{weiner99} and references therein).
 
Another aspect of the theoretical developments in the cold-molecule field, 
where the relative motion of two atoms has to be accurately described from 
a distance of around one to thousands of atomic units,   is the implementation  of numerical 
grid methods to represent the continuum wave function in the ground state and 
the vibrational wavefunction in the excited state, with grid steps adapted to 
the local de Broglie wavelength: such methods are routinely used in at least six 
papers in this volume.

\subsection{ Making cold molecules from cold atoms via magnetic Feshbach Resonances}
\label{ssec:feshbach}

The possibility of creating ultracold molecular Bose-Einstein condensates
via resonant collisions 
\cite{bill_stwalley,eite1,eite2,eite3,mitja,mitja2,mitja3,fesh1,fesh2,fesh3,fesh4,fesh5,fesh6}
became a reality in 2003 
\cite{grj03,jba03,zss03,bkc04}. The idea is simple but its realization is not, as a very
dense gas of quantum-degenerate (or near-degenerate) precursor atoms is
required. Fortunately, technology has advanced enough so that ultracold
alkali metal atom gases are now routinely produced by more than 30 groups
worldwide.


In order to stitch ultracold atoms together and create molecules, a Feshbach
resonance can be used. The reader is directed to other publications 
(\cite{kleppPT04} and references therein)
 for a more detailed picture. In brief, in the ultracold regime
the collision of two atoms is described by the scattering length. This
scattering length is very sensitive to the potential of the interacting
atomic pair (molecular potential). If the atoms or the molecule or both are
paramagnetic, the potential can be tuned by an external magnetic field. A
Feshbach resonance occurs when the energy of the unbound state (the atomic
pair) and of the bound state (the diatomic molecule) become equal. 
The scattering length changes at a Feshbach resonance. 
It can be made positive (corresponding
to a repulsive interaction) or negative (attractive interaction) and its
magnitude arbitrarily large (within the limits of unitarity). By tuning
through the resonance the atom pair can be driven into the bound molecular
state. The binding energy of the molecule is tiny, typically in the
mK range. 
Molecules created from bosonic atoms are quickly destroyed 
by collisions, while molecules produced from fermionic atoms may exist for 
a long time \cite{mitja,mitja2,mitja3}.
It is believed that
it might be possible to transfer them to more tightly bound states using,
for example, pulsed Raman laser fields \cite{koch2004} (see also the discussion 
in Sec. 2.1). At the workshop, this approach to
producing heteronuclear molecules was discussed but no experimental results
were then available. Now there are at least two groups that have found
Feshbach resonances in heteronuclear systems \cite{szs04,igo04}. This
greatly raised the expectation that polar molecules will be produced in this
way.

The rich resonant structure of molecular spectra observed in ultracold collision 
experiments cannot be explained without a rigorous theoretical analysis. 
{\bf Bhattacharya, 
Baksmaty, Weiss,} and  {\bf Bigelow} (pp xx-yy)
present an article in this issue with a clear discussion 
of methods for calculating Feshbach resonances in heteronuclear alkali dimers. 
The presentation is both simple enough to be understood by a general reader 
and comprehensive enough to be of interest to an experimentalist.

\vspace{0.03in}
\section{Methods for cooling, slowing, and trapping polar molecules}

(a) \emph{Buffer-gas cooling }is a versatile technique which provides the
means to cool any molecules (or atoms) to temperatures of a few hundred
millikelvin. This temperature is determined by the equilibrium vapor
pressure of the He buffer gas that is used to thermalize the molecules via
elastic collisions. The technique, pioneered by the Doyle group in 1997 \cite
{eu}, has been used in a number of studies since to cool paramagnetic atoms
or molecules and to load them into a magnetic trap \cite{primer}. A
recalcitrant problem has been the introduction of the molecules to be cooled
into the cryogenic buffer gas. Solving this problem has been of primary
importance, since the number of molecules that can be cooled and trapped is
limited mainly by the number of molecules that can be brought into the
buffer gas \cite{cahnature}. The techniques that have been applied so far,
laser ablation \cite{primer} and capillary filling \cite{capillary}, suffer
from serious drawbacks. Capillary filling is suitable only for stable
molecules that remain gaseous at low temperatures. Laser ablation offers
more versatility (including in situ production of unstable species), but the
yield and variety of molecules produced from a solid precursor and, indeed, the
precursor's very chemical identity are hard to predict.

The paper by \textbf{Egorov, Campbell, Friedrich, Maxwell, Tsikata, van
Buuren, }and\textbf{\ Doyle} (pp. xx-yy) describes a new technique of
injecting molecules into the buffer gas, one which offers significant
advantages over the existing techniques. The reported technique makes use of
a molecular beam made up of the species of interest and employs the beam to
deliver this species into the cryogenic cell. The work demonstrates that a
subtle balance can be struck between the molecular beam flux into the cell
and the flux of the He buffer gas out of the cell such that the injection
and thermalization occur on time scales shorter than the diffusion loss
time. The beam-loaded buffer gas cooling technique is demonstrated with a
pulsed beam of NH radicals produced from NH$_{3}$ in a glow discharge. The
maximum number of ground-state (i.e., thermalized) NH$(^{3}\Sigma ^{-},$v$%
=0,N=0)$ molecules loaded into the cell of volume of $150$ cm$^{3}$ was
measured to be 10$^{12}$. The radicals were detected via laser-induced
fluorescence. In addition, the authors make the point that the molecular
beam can be rid of unwanted species by applying standard electrostatic or
magnetic filters.

(b) \emph{Stark acceleration or deceleration }provides the means to alter
the velocity of polar molecules at will and to simultaneously select the
molecules' internal states. Although conceived almost half a century ago,
the technique was implemented only recently, in 1999, by the group of
Meijer \cite{bbmprl1999}. The technique relies on time-dependent
inhomogeneous fields generated by an array of field stages, to repetitively
alter the Stark potential energy of the molecules and, by energy
conservation, also their kinetic energy. In a Stark decelerator, the
potential energy is increased at each decelerator stage and thus the kinetic
energy is correspondingly reduced (and vice versa in an accelerator). In
order to accelerate/decelerate a bunch of molecules with a finite
distribution of positions and velocities, the acceleration/deceleration
process must take place under the condition of phase stability, much
appreciated in charged-particle accelerator physics 
\cite{veksler,mcmillan}. Stark deceleration was first demonstrated with an array of $63$
pulsed electric field stages. This was shown to decelerate metastable CO
molecules from $225$ m/s to $98$ m/s \cite{bbmprl1999}.

Since 1999, the Stark decelerator underwent many refinements and was
combined with other devices \cite{pra_of_meijer}, most notably a trap \cite{trapnature}, a
buncher, and a storage ring \cite{ringnature} (see below). Also other
laboratories have since implemented Stark deceleration 
\cite{tbh04,hgould2,ye,rydberg1,rydberg2,rydberg3}.

\textbf{Friedrich}'s paper (pp. xx-yy) provides a rigorous description of
the dynamics of molecules in a Stark accelerator/decelerator. The paper
opens with a Fourier analysis of the time-dependent inhomogeneous Stark
field, of the kind generated in current experiments by linear switchable
field arrays \cite{oldmodel}. This is used as a basis for developing a
quasi-analytic model of a Stark accelerator/decelerator. This model depends
on just a small number of parameters; the lowest-order version (referred to
as the first-harmonic accelerator/decelerator model) depends on a single
Fourier coefficient, one which accounts for the spatial dependence of the
inhomogeneous electric field. Yet, the simplest version of the model is
shown to explain the key features of the accelerator/decelerator (such as
phase stability and longitudinal acceptance), and to capture the dynamics of
the acceleration/deceleration process semi-quantitatively. An extensive
discussion of the dynamics exploits the isomorphism of the first-harmonic
accelerator/decelerator with a biased pendulum. Also, the general properties
of the velocity of molecules in a phase-stable accelerator/decelerator are
examined, and found to reveal a similarity between a Stark
accelerator/decelerator and a flying accordion.

The paper by \textbf{van Veldhoven, K\"{u}pper, Bethlem, Sartakov, van Roij,
}and\textbf{\ Meijer} (pp. xx-yy), demonstrates the benefits of using slow
molecules in microwave-UV double-resonance spectroscopy. The resolution of
the inversion tunneling spectrum of $^{15}$ND$_{3}$ is shown to be boosted
by Stark-decelerating the molecules to a velocity of $52$ m/s. The
transit-time broadening is thereby reduced by a factor of $10$ compared
with undecelerated molecules moving at $280$ m/s. This results in a fully
resolved hyperfine spectrum with a $1$ kHz linewidth (the spectrum was taken
for the $J=1,K=1$ state over $1.43$ GHz). A simulation of the measured
spectrum, based on a generalized tensor coupling scheme, is also described.
As the authors point out, the inversion frequency in ammonia depends
sensitively on the electron/proton mass ratio, and could be used to monitor
the ratio's possible time dependence. In order for the techniques to become
competitive \cite{astrotimedep}, the present accuracy needs to be further
improved, by a factor of $10^{6}$. This should become possible once an
ammonia molecular fountain is implemented, allowing for an interaction time of
as much as a second. We note that deuterated ammonia molecules, due to their
small inversion splitting, exhibit a first-order Stark effect at `all' field
strengths (including the low ones). This makes them a favorite choice for
studies on molecular manipulation.

The paper by \textbf{Hudson, Bochinski, Lewandowski, Sawyer, }and\textbf{\ Ye%
} (pp. xx-yy) analyzes the experimental conditions needed for an efficient
Stark deceleration of a beam of OH radicals. Calculations are presented that
show the fraction of molecules decelerated, given a certain emittance of the
molecular beam and acceptance of the decelerator. The calculated fraction is
compared with experimental data, and a good agreement is found. The fraction
decreases with the decreasing goal velocity and drops to about $0.1$\% for a
goal velocity of about $30$ m/s, for molecules moving initially at $370$
m/s, and originating from a hot beam (whose mean velocity and velocity width
are, respectively, $415$ m/s and $90$ m/s). The paper makes the point that
the fraction of molecules decelerated could be substantially increased if a
cold source were used.

(c) A \emph{storage ring} offers an alternative to trapping. Unlike a trap,
a storage ring can accommodate molecules whose longitudinal velocities are
not all that low. However, since the ring radius scales with the square of
the velocity, table-top storage rings require samples of molecules slowed
down to velocities on the order of $10$ m/s. A prototype storage ring for
low-field seeking states of polar molecules was implemented by Crompvoets 
{\it et al.} \cite{ringnature} and used to store about 10$^{6}$ ND$_{3}$ molecules
for up to $430$ ms corresponding to $50$ round trips (the storage time was
limited by background gas scattering). The storage ring of a $25$ cm
diameter was made up of a hexapole lens bent onto itself. Before loading,
the molecules were Stark-decelerated to a velocity of $92$ m/s. In addition,
a buncher was implemented, to narrow the longitudinal velocity spread of the
decelerated molecules to about $300$ $\mu $K.

\textbf{Nishimura, Lambertson, Kalnins, }and\textbf{\ Gould }(pp. xx-yy)
propose a storage ring for high-field seekers. The authors envision a
storage ring $3$ m in diameter, consisting of eight octants. Each octant
contains a buncher and a pair of alternating-gradient focusing triplet
lenses. In order to obtain net focusing in both transverse directions, the
lenses are arranged in a sequence with gradients alternating in sign (the so
called alternating-gradient focusing \cite{auerbach}). Simulations indicate
that such a ring would store, e.g., ground-state CH$_{3}$F molecules with a
longitudinal kinetic energy of $2$ K for several minutes. The molecules
would be produced in a pulsed supersonic expansion and Stark decelerated
before injection into the ring. For a peak beam intensity of $10^{19}$ sr$%
^{-1}$ s$^{-1}$ and a nearly perfect phase-space matching between the beam,
decelerator and ring, each bunch in the ring would consist of about $10^{8}$
molecules and would have a density on the order of $10^{9}$ cm$^{-3}$ in the
straight sections throughout the storage time. Such a high density may
foster evaporative cooling. Apart from storing singular bunches, the ring
may be able to hold up to $200$ bunches simultaneously. This would provide a
large flux desirable, for instance, for collisional or spectroscopic
experiments. These could be carried out without field perturbations, in the
field-free regions of the storage ring.\emph{\ }

(d) The paper by\textbf{\ Junglen, Rieger, Rangwala, Pinkse, }and\textbf{\
Rempe} (pp. xx-yy) describes an \emph{electrostatic guide for slow polar
molecules in low-field seeking states}, which shares some of its key
features with the storage ring of Crompvoets {\it et al.} \cite{ringnature}. A
guide for molecules may find use in molecular interferometry and in the
study of slow collisions.

The technique selects the low-velocity tail of a Maxwell-Boltzmann velocity
distribution present in a thermal effusive molecular beam. Transverse
velocity selection is achieved by injecting the molecular beam into a
quadrupole electrostatic field generated by rod electrodes aligned along the
beam axis. By bending the rod electrodes, also longitudinal velocity
filtering is achieved. The technique was demonstrated with ND$_{3}$
molecules. A flux of $10^{10}$ translationally cold ND$_{3}$ molecules per
second could be achieved. The longitudinal temperature of the molecules was
measured, via time-of-flight mass spectrometry, to be about $4$ K. The
efficiency of the technique benefits from cooling the beam source. However,
the effusive flow regime needs to be maintained, in order for the
hydrodynamic velocity to remain negligible and the low-velocity tail in
place. This of course precludes cooling of the internal degrees of freedom
of the molecules. Although the Stark effect is somewhat greater for low
rotational states, these are not selected efficiently. As a result, the
internal temperature of the translationally cold molecules picked up by the
Stark field remains close to that of the source.

(e) The Earnshaw theorem precludes trapping of atoms or molecules via their
high-field seeking states in \emph{static} 
electric or magnetic fields \cite
{auerbach}. Since ground states of atoms and molecules are always high-field
seeking, there has been a considerable effort to find ways of circumventing
Earnshaw's curse, by devising traps for high-field seekers made up of
time-dependent electromagnetic fields. These comprise switched static 
fields \cite{piek} 
(ac traps) or radiative fields, both resonant (MOT's) and nonresonant
(optical dipole traps) \cite{jba03,mch93}.
 A common drawback of these traps is their small
volume and depth; these significantly limit the ability to apply evaporative
cooling which could otherwise be used to drive the trapped molecules into
the ultracold regime ($\leq 1$ mK).

\textbf{DeMille}, \textbf{Glenn}, and \textbf{Petricka} (pp. xx-yy) propose
a deep, large-volume \emph{microwave trap for high-field seeking states} of
cold/slow polar molecules. A maximum of free-space electric field strength
is produced in a confocal cavity, pumped by $2$ kW of resonantly coupled
microwave power. Field strengths on the order of $10$ kV/cm in the TEM$_{00}$
mode spread over a volume of 1 cm$^{3}$ seem feasible. The trapping is based
on a near-resonant electric-dipole interaction of the microwave field with
the rotational ground state of a polar molecule. The interaction couples the
ground state to higher rotational states and modifies its eigenenergy. For a
red detuning of the microwave field, the interaction is attractive and so
the eigenenergy drops with respect to the field-free value. For a linear
polarization of the microwave field, multiphoton transitions complicate the
resulting energy-level pattern. However, for a circular polarization (and
red detuning), multiphoton transitions can no longer occur, and the energy
level pattern greatly simplifies. In particular, the ground-state level
becomes free of avoided crossings with other levels and its dependence on the
strength of the microwave field becomes similar to that of a ground state
polar molecule in a DC field. Thus, the ground-state eigenenergy of a polar
molecule in a red-detuned circularly polarized microwave field can be used
for stable trapping. Because of the long radiative lifetimes of the excited
rotational states, the red detuning can be minimized, without the danger of
heating the trapped molecules by photon scattering (which bedevils trapping
at optical frequencies). This maximizes the drop of the eigenenergy with
field strength and thus the trap depth. Indeed, according to the
dressed-state analysis presented in the paper, the trap depth for the
molecular ground state scales roughly as $\frac{1}{2}\mu \varepsilon _{0}$,
where $\mu $ is the body-fixed electric dipole moment of the molecule and $%
\varepsilon _{0}$ the effective field strength at the center of the trap.
Hence for $\mu \approx 1$ Debye and $\varepsilon _{0}\approx 10$ kV/cm, the
expected trap depth is on the order of $100$ mK.

The paper by DeMille {\it et al.} also discusses various options of accumulating
and cooling the molecules in the microwave trap. Among these, the
possibility of loading multiple pulses of molecules in a low-field seeking
state and pumping them into the trappable ground state seems particularly
appealing (a similar scheme has been proposed earlier by van der Meerakker
{\it et al.} \cite{pumpload}). Another cooling scheme envisioned is evaporative
cooling driven by elastic collisions. The elastic collision cross sections
are shown to be dramatically enhanced by the alignment of the molecular
dipoles by the trapping field and the ensuing strong dipolar interaction.
Elastic cross sections for collisions between ground state SrO molecules are
estimated to come close to $10^{7}$ \AA $^{2}$. This would suffice for a
thermalization of the trapped molecules at number densities as low as 10$%
^{8} $ cm$^{-3}$.

(f) \emph{Single-collision slowing in crossed molecular beams.} The paper by 
\textbf{Elioff, Valentini, }and\textbf{\ Chandler} (pp. xx-yy) describes an
experiment that demonstrates a rather general technique of producing slow
molecules, by billiard-like collisions in crossed molecular beams.
Significant numbers of slow NO molecules ($\approx 15$ m/s, corresponding to
a translational temperature of about $0.5$ K) in a single internal state ($%
^{2}\Delta _{1/2},$ v$=0$, $J=15/2$) are generated by scattering of a pulsed
molecular beam of NO by a beam of Ar. The number of slow molecules produced (%
$\approx 10^{5}$ per pulse) corresponds to a fraction of about $10^{-5}$ of
the overall number of molecules that are colliding during a beam pulse. The
technique relies on the cancellation of the laboratory velocity of the
scattered molecules. This occurs when the laboratory velocity of the center
of mass of the colliding partners (NO + Ar) becomes equal in magnitude but
opposite in direction to the center-of-mass recoil velocity of the scattered
molecule (NO). The velocity spread of the molecules that satisfy this
cancellation condition depends on the velocity spread of the NO beam. For
the scattering of partners with the same mass, the condition is fulfilled
for elastic collisions, in which case, interestingly, the energy spread of
the slow molecules is found to decrease to zero. The technique has been
demonstrated using a state-of-the-art detection scheme based on
state-specific REMPI ionization and micro-channel plate imaging of the NO$%
^{+}$ ions produced. 

(g) A class of atoms, best represented by the alkalis, can be efficiently
cooled by repetitive absorption of photons that counter-propagate with
respect to the atoms (Doppler cooling). Since, in the visible range, the
momentum transferred by a single photon is small, typically thousands of
such photons need to be absorbed before a thermal alkali atom can be brought
to a halt. In order for the absorption to be repetitive, the re-emission of
the absorbed photon must send the atom into the same initial state where it
started. Hence we speak of a closed absorption/emission cycle. Finding
molecules with a closed absorption/emission cycle is difficult, if not
impossible. As a way out, Bahns {\it et al.} have proposed a sequential scheme
which uses an array of stimulated Raman sidebands to repump molecules that
have fallen out from the cooling cycle \cite{bsg96}.

\textbf{Di Rosa}'s paper (pp. xx-yy) examines whether a certain class of%
\emph{\ }molecules\emph{\ }can be\emph{\ laser-cooled }without extensive
repumping and, provided the molecules are paramagnetic, whether they can be%
\emph{\ trapped in a magneto-optical trap} (MOT). The paper identifies the
criteria that molecules must meet in order to be amenable to Doppler cooling
and provides a list of such molecules. The criteria include a strong
vibronic transition, with a near-unit Franck-Condon factor, and no
intervening electronic transitions that could divert photons from a closed
cooling absorption/emission cycle. Out of a list of ten diatomic molecular
candidates (six of which are paramagnetic), Di Rosa chose 
CaH(A$^{2}\Pi_{1/2}$-X$^{2}\Sigma^+$) as a representative example.

The number of absorption/emission cycles needed to decelerate a molecule of
mass $m$ from an initial velocity v to a standstill with radiation of
wavelength $\lambda $ is $N=m$v$\lambda /h$, which comes close to $4500$ for
CaH(A,$v^{\prime }=0$-X,$v^{\prime \prime }=0$) moving initially at v$%
\approx 100$ m/s. For a cycling transition with a Franck-Condon factor of $%
0.9995$, the probability, $p$, of spontaneous emission to levels outside of
the absorption/emission cycle is then $p=0.0005$. Hence the probability that
a molecule will remain in the absorption/emission cycle is $(1-p)^{N}\approx
0.1$ and so about $10\%$ of the molecules will be cooled. Although a
Franck-Condon factor of about $0.9995$ is quite rare, in some molecules,
such as CaH(A-X), the Franck-Condon factors for two neighboring levels, $%
v^{\prime }=0,1\rightarrow v^{\prime \prime }=0$, add up to a value close to
$0.9995$. This would then require a single repumping laser, to drive the $%
v^{\prime }=1\rightarrow v^{\prime \prime }=0$ transition.

The experimental work reported by Di Rosa 
 is a first step towards a CaH MOT. It examines
experimentally a laser-cooling cycle for CaH, in particular the underlying
hyperfine structure of the A-X band. 
He estimates that
the lifetime in a CaH MOT will
be about $1$ ms. At a thermal loading rate of the CaH vapor, the MOT could
hold about $10^{5}$ molecules.

(h) The work of \textbf{Bertelsen, Vogelius, Jorgensen, Kosloff,} and {\bf Drewsen}
(pp. xx-yy) takes advantage of the low temperatures ($\approx 10$ mK)
available in Coulomb crystals to control the photodissociation dynamics of
MgH$^{+}$ ions embedded in the crystal. The molecular ions are
sympathetically cooled into the crystal by laser-cooled atomic ions. The
authors foresee that the cold crystal environment along with the tight
spatial localization ($\approx 1$ $\mu $m) will make it possible to control
the branching ratio between the Mg$^{+}$(3p)+H and Mg(3s$^{2}$)+H$^{+}$
dissociation channels, by radiative coupling of MgH$^{+}$(A) to its B and C
electronic states. This will be effected by two independent laser fields.
The paper presents calculations which indicate that the branching ratio
could be varied at will, either by varying the intensity of the lasers
involved or the frequency of one of the lasers.

(i) Another technique for producing cold molecules relies on thermalization 
in the cryogenic environment provided by helium nanodroplets. This technique, 
pioneered in 1992 ~\cite{goyal92}, has been implemented by several groups and 
used to improve the resolution of molecular spectroscopy \cite{dong,lehnig},
study nanoscopic superfluidity \cite{grebenev,toennies}
and to foster 
self-assembly of non-equilibrium structures 
\cite{burnham},
 to name just a few.

He nanodroplets are produced by a supersonic expansion of He at high 
pressures (5-45 bar) and low temperatures (5-25 K). The nanodroplets cool 
down by evaporation to a temperature of 380 mK or 150 mK if $^4$He
or $^3$He are used, respectively; the uniform temperature of the nanodroplets
is determined essentially by their surface tension.
The resulting size distribution is log-normal, 
peaking typically at thousands of He atoms.

The nanodroplets can be doped by atoms or molecules, by passing the 
nanodroplets beam through a vapor pick-up cell. The atoms or molecules 
quickly thermalize, in all their degrees of freedom, with the cold He 
nano-environment. They can recombine with other dopant atoms or molecules, 
if these are picked up by the same nanodroplet (the pick-up statistics is 
Poissonian); the binding energy is released into the He nano-bath. The 
ensuing species can then be sensitively probed via high-resolution spectroscopy.
Atoms in He nanodroplets can thus form dimers (or larger oligomers).

The formation of cold heteronuclear alkali 
dimers NaK in helium droplets was previously 
observed  by Higgins {\it et al.} \cite{higgins98,higgins2000}. 
In the present issue, the experimental paper by 
{\bf Mudrich, B\"unerman, Stienkemeier, Dulieu} and {\bf Weidem\"uller}   
(pp xx-yy)
reports the observation of the efficient  formation of LiCs and NaCs
in helium nanodroplets.
  The molecules are 
excited by a laser and then photo-ionized. 
The paper reports the excitation 
spectra recorded from mass selective photo-ionization. 
The potential curves correlating with the Li(2s) + Cs(5d) and Na(3s) + Cs(5d) 
asymptotic limits are obtained based on the measurement and compared with theoretical results from
{\it ab initio} calculations with perturbative treatment of the fine structure. 
The discussion shows the need for a global frequency shift of the computed potentials. 
 We note that little was
known about the spectroscopy of LiCs and NaCs in the triplet state before
this study was undertaken.

The authors point to the intriguing possibility of decelerating and
trapping nanodroplets - along with the dopant molecules. This, they
envision, could be achieved by first ionizing the nanodroplets (by
electron impact or capture) and then applying a stopping potential. The
cold and slow molecules could then be photo-desorbed and confined in an
electrostatic trap. The flux of the molecule-doped He nanodroplets is
estimated to be about $10^{10}$ s$^{-1}$.


\noindent%
\section{Theory of cold collisions involving molecules}
The production of ultracold molecules has led chemists to ask the questions:
What is the efficiency of chemical reactions and inelastic energy transfer
in molecular collisions at low temperatures? Are chemical reactions possible
at temperatures near absolute zero?

Based on the analysis of Landau and Lifshitz \cite{landau}, Balakrishnan,
Kharchenko, Forrey and Dalgarno showed that elastic and inelastic molecular
scattering can be characterized in the limit of ultralow collision velocity
by a complex scattering length \cite{bala1}

\begin{equation}
a_{n} = \alpha_{n} - i \beta_{n}.  \label{sl}
\end{equation}

\noindent%
The imaginary part of the scattering length $\beta _{n}$ is directly
proportional to the total cross section for inelastic scattering in the
state $n$

\begin{equation}
\beta_{n} = k_{n}\sigma_{n}^{\mathrm{in}}/4\pi,  \label{beta}
\end{equation}

\noindent%
where $\sigma _{n}^{\mathrm{in}}=\sum_{n^{\prime }\ne n}\sigma
_{n\rightarrow n^{\prime }}$, $k_{n}=\mu v_{n}/\hbar $, $v_{n}$ is the
collision velocity and $\mu $ is the reduced mass of the collision system.
Eq. (\ref{beta}) can be used to evaluate the zero temperature rate
coefficient for inelastic energy transfer or chemical reaction

\begin{equation}
R_{n}(T = 0 \hspace{0.1cm} \mathrm{K}) = 4 \pi \beta_{n}\hbar/\mu,
\label{zerorate}
\end{equation}

\noindent%
Because $\sigma _{n}^{\mathrm{in}}\sim 1/k_{n}$ according to the Wigner law,
$\beta _{n}$ is constant in the limit $k_{n}\rightarrow 0$ and the rate
coefficient (\ref{zerorate}) is finite at zero temperature.

The elastic cross section is given by

\begin{equation}
\sigma_{n \rightarrow n} = 4 \pi (\alpha_{n}^2 + \beta_{n}^2).  \label{alpha}
\end{equation}

The positions and lifetimes of the bound levels of the collision complex
lying just below dissociation threshold can be estimated from the real and
imaginary parts of the scattering length as follows \cite{bala1}:

\begin{equation}
E = - \frac{\hbar^2}{2\mu|a_{n}|}(\cos{2\gamma_n} + \mathrm{i} \sin{2\gamma_n%
}) = E_n - \mathrm{i} \Gamma_n/2,  \label{res}
\end{equation}

\noindent
where $\gamma _{n}=\arctan {(\beta _{n}/\alpha _{n})}$. The energy $E$ is
real for single channel scattering because $\beta _{n}=0$ and $\gamma _{n}=0$. 
When there is more than one open channel, the bound levels can decay
through transitions into lower levels with a lifetime

\begin{equation}
\tau_n = \hbar/\Gamma_n = \mu |a_n|^4/(2\hbar\alpha_n\beta_n)  \label{life}
\end{equation}

\noindent%
that can be rewritten in terms of the zero energy elastic cross section and
zero temperature rate coefficient (\ref{zerorate}) as follows:

\begin{equation}
\tau _{n}=\frac{\sigma _{n\rightarrow n}(k_{n}\rightarrow 0)|a_{n}|^{2}}{%
2R_{n}(T=0)\alpha _{n}}  \label{lifer}
\end{equation}

\noindent%
Forrey \textit{et al.} \cite{bala3} showed that the expression (\ref{lifer})
can be used to estimate the positions and lifetimes of Feshbach resonances
occurring in the incident channel in $s$-wave collisions. The derivation of
Eq. (\ref{lifer}) is based on multichannel effective-range theory and it can
be applied only to the states that lie close to thresholds. More details on
threshold phenomena in atom-atom, atom-molecule and electron-atom collisions
can be found in a comprehensive review by Sadeghpour \textit{et al.} \cite
{hossein}, the review by Krems \cite{roman_review} and the paper of Mies and
Raoult \cite{mies}.

The Wigner law establishes that rate coefficients for inelastic collisions
and chemical reactions at zero temperature are finite. The magnitudes of the
zero temperature rates for inelastic scattering of atoms and molecules were,
however, not known until recently. The first calculations of rate constants
for vibrational relaxation in atom - molecule collisions at ultralow
temperatures were carried out by Schwenke and Truhlar \cite{schwenke}.
They demonstrated that conventional quantum theory of
vibrational relaxation is adequate for interpretation of low energy atom -
molecule collisions and showed that low energy cross sections follow the
trends predicted by Wigner. The rate constants for the He - I$_2$ system
studied by Schwenke and Truhlar were, however, extremely small in the zero
temperature limit and it was concluded that the zero energy divergence of
inelastic cross sections may not affect the rate coefficients for inelastic
scattering at temperatures $\ge 10^{-4}$ K in molecular systems. More recent
calculations demonstrate that often this is not so.

The studies of the He + H$_2$ and other collision systems 
\cite{bala4,bala5,bala6,bala7,bala8,bob1,bodo1,roman1,roman2,siska,roman3,bob2,bob3} 
showed that zero temperature rate coefficients for vibrational, rotational and
electronic energy transfer in atom - atom and atom - molecule collisions may
have substantial magnitudes. It was demonstrated by several authors that the
attractive part of the interaction potential in the entrance collision
channel is critical for the dynamics of ultracold scattering. Dashevskaya
and Nikitin presented a simple explanation of this phenomenon using a
generalized Landau-Lifshitz treatment of a collision problem \cite{nikitin}.

The stability of rotationally hot molecules in a cold buffer gas has been 
explored in rigorous calculations of atom - molecule collisions at ultracold
temperatures presented in this issue (\textbf{Forrey} pp. xx-yy).
Interesting implications for observation of ultracold super-rotors and
quasiresonant energy transfer at very low collision energy should be
appealing to experimental investigators.

Chemical reactions can be of two types: insertion reactions and abstraction
reactions. The insertion reactions are often barrierless. 
Quantum mechanical calculations of chemical reactions are usually 
carried out in hyperspherical coordinates.
Probabilities of chemical reactions are obtained 
from the solution of 
coupled differential equations on a grid of a single 
coordinate - the hyperradius. The total
wave function of the reactive complex may be expanded 
in terms of products of asymptotic functions corresponding to different 
reaction arrangements or in a basis of functions that vary with 
the hyperradius. The former procedure is suitable for the description 
of abstraction reactions, while the latter should be used for 
insertion reactions. 

An example of an
insertion reaction is the reaction of an alkali atom with an alkali diatomic
molecule, like Na + Na$_2$. 
Sold\'{a}n and coworkers used a hyperspherical
coordinate quantum mechanical method 
of Launay and Le Dourneuf \cite{launay}
to investigate the Na + Na$_2$ reaction
at ultracold temperatures \cite{jeremy}. They showed that the cross sections
for the reaction removing the vibrationally excited Na$_2(v=1)$ molecules
follow the Wigner law and the rate coefficient for the reaction is large in
the limit of zero temperature. Recently, the same authors have extended their 
calculation to study the Li + Li$_2$ reaction at zero temperature 
\cite{jeremy_condmat}. It was shown that
for low-lying bound states of Li$_2$ there 
is no systematic difference between the inelastic collision rates for 
molecules formed from fermionic and bosonic Li.

The abstraction reactions have a potential barrier which separates the
reactants from products. An example of an abstraction reaction is the
chemical reaction of F with H$_{2}$. Because the reaction barrier always has
a large magnitude on the ultracold energy scale, it might be expected that
the abstraction reactions would not occur at ultracold temperatures.
Balakrishnan and Dalgarno presented the first investigation of an
abstraction reaction at ultracold temperatures \cite{bala_c1}. It was found
that the cross sections for the chemical reaction F + H$_{2}$ $\rightarrow $
FH + H follow the $s$-wave Wigner law and have surprisingly large magnitudes
at ultralow energies. The rate coefficient for the reaction at zero
temperature was calculated to be $1.25{\times }10^{-12}$ cm$^{3}$ s$^{-1}$.
It was suggested that the large rate of the reaction at ultracold
temperatures is due to tunneling of the light H atom under the reaction
barrier. Several later studies of F + D$_{2}$ collisions 
\cite{bodo_c1,bodo_c2,bala_r1} indicated that tunneling plays an important role in the
reaction. A very recent article by Bodo, Gianturco, Balakrishnan and
Dalgarno demonstrated that the chemical reaction of F with H$_{2}$ at
ultracold temperatures is assisted by a virtual resonance state near
collision threshold \cite{bodo_c3}. A slight variation of the mass of H$_{2}$ shifts the
virtual level in and out of resonance with the collision energy, leading to
dramatic enhancement or suppression of the reaction rate.

Two articles in this issue (\textbf{Weck} and \textbf{Balakrishnan} pp. xx-yy and
\textbf{Bodo} and \textbf{Gianturco} pp. xx-yy) report calculations of chemical
reactions with large activation barriers in the limit of zero temperature.
It is demonstrated that abstraction chemical reactions may be very fast at
ultracold temperatures and the reactivity is enhanced to a great extent by
internal excitation of the reactants. In particular, it is shown that the
rate of the F + H$_2$ reaction is larger than the rate of collisional
relaxation of the reactants when the H$_2$ molecule is excited to an energy
level above the activation barrier. These studies will potentially lead to
the development of the field of ultracold chemistry as they show that
chemical reactions do occur at zero absolute temperature. Studies of
ultracold chemistry in external fields will open up the field of controlled
chemistry.

The possibility of the creation of ultracold molecules is determined by
their lifetime in an external field trap. The stability of cold and
ultracold molecules in magnetic traps was studied in a series of papers by
Bohn and coworkers \cite{bohn1,bohn2,bohn3,bohn4} and Krems and coworkers
\cite{krems1,krems2,krems3}. Krems and Dalgarno proposed a theory for
quantum mechanical calculations of molecular collisions in a magnetic trap
and showed that trap loss of magnetically trapped diatomic molecules in
electronic $^2\Sigma$ or $^3\Sigma$ states and in the rotational ground state
is determined by the coupling to rotationally excited states \cite{krems1}.
The rotationally excited states are not energetically allowed at low
collision energies so collision-induced spin relaxation occurs through
virtual excitations in the reactive complex and it is slow. This study
identified the range of molecules that may be amenable to magnetic trapping
and evaporative cooling to ultracold temperatures. Novel states in dimers of
ultracold molecules linked by an electric field were found by Bohn and
coworkers \cite{bohn5}.

To explore the prospects for sympathetic cooling of molecules by collisions
with ultracold atoms, Sold\'{a}n and Hutson computed highly accurate
interaction potentials of the Rb--NH complex in various spin states 
\cite{jeremy2}. They showed that collision dynamics of ultracold Rb atoms with NH
molecules may be complicated by the coupling to ion-pair states leading to
dramatic enhancement of the interaction strength.

Krems and Dalgarno derived threshold laws for reorientation of electronic
angular momentum in atomic and molecular collisions in the absence of
external fields \cite{roman_threshold}. Together with the computational
study of Volpi and Bohn \cite{bohn4}, this analysis indicates that the
Zeeman relaxation rate in ultracold collisions of atoms or molecules should
increase dramatically with increasing field strength.

An important discovery is reported in the article of this issue 
(\textbf{K{\l}os, Rode, Rode, Cha{\l}asi{\'n}ski,} and \textbf{Szcz\c{e}\'sniak} pp. xx-yy)
describing interactions of non-$S$-state transition metal atoms with He. 
Non-$S$-state atoms interact upon collisions with He effectively like diatomic
molecules \cite{aquil,roman_jpc} and collisional cooling of non-$S$-state
atoms in a magnetic trap has been predicted to be impossible \cite
{roman_shaped,santra}. The study of K{\l}os and coworkers, however,
demonstrates that the angular dependence of the interaction potential
between transition metal atoms and He is dramatically suppressed. This
indicates that transition metal atoms should be amenable to buffer-gas
loading in a magnetic trap and, possibly, evaporative cooling to ultracold
temperatures. Trapping of non-$S$-state atoms will greatly expand the scope
of ultracold molecular physics. Photoassociation of ultracold non-$S$-state
atoms will produce ultracold non-$\Sigma$-state molecules.


\section{Conclusions and future directions}
\label{sec:conclu}

We hope that the readers of this issue will be  convinced of the rapid expansion of the field of ultracold molecules. We should note the combination of traditional physics (accurate molecular spectroscopy) and 
very recent research advances (molecular quantum gases). We should also note the strong interplay between theory and experiment, which appears in many papers of this issue.
 We expect the next few years will be marked
by the development of improved techniques for the production of ultracold
molecules, and by more fundamental studies with cold and ultracold molecules
as model systems and as a tool for very precise measurements.

It is clear from the present series of papers 
that there are many roads to cold molecules.
The groups working with indirect methods like photoassociation of ultracold 
atoms will 
develop efficient ways of creating molecules in the vibrational ground level, 
making use of various technologies (e.g., Raman schemes, chirped pulses) 
that have yet to be fully studied. 
The groups working with direct methods will develop techniques to reach lower temperatures. 
In either case, trapping molecules and detecting them is an important issue.

The development of \emph{new types of traps} for molecules in
\emph{high-field seeking states} will likely be pursued as a priority by
several groups. Apart from the microwave trap proposed in this issue, it
is the ac trap for polar - and perhaps even paramagnetic - molecules that
raises great expectations. Such traps could confine ground state
molecules, which are immune to dipolar relaxation and other loss
processes. This would greatly improve the chances for achieving
\emph{evaporative cooling} - and reaching the ultracold regime. Such traps
offer a large volume and might be loaded by optical pumping of slow
molecules in suitable excited states. The presence of the trapping field
is expected to greatly enhance the scattering cross section by aligning
the molecular dipoles, and thus reduce the requisite number density of the
trapped molecules. Alternatively, the critical density conducive to
evaporation may be achieved by bunching.  Evaporative cooling is also
likely to be attained in magnetic traps, which can be loaded with
large numbers of molecules using the beam-loaded buffer gas technique.
Another possibility is to use a hydride molecule plus laser cooling to
directly load the molecules into a magnetic, electrostatic or ac
trap. \emph{Sympathetic cooling} of molecules (or molecular ions) with
laser cooled atoms (or atomic ions) is also being pursued in a number of
laboratories. 

Cooling Rydberg molecules is an interesting challenge to explore
\cite{rydberg1,rydberg2,rydberg3}.

During the next few years, new experiments are likely to be
carried out that will \emph{make use of cold molecules}. Scattering of
slow molecules by atoms confined in a MOT, or of successive bunches in a
decelerator, storage ring or trap by one another are among the pursuits
likely to be taken up in several laboratories. The implementation of a
molecular fountain is a prominent goal.

Chemical reactions have been shown to occur rapidly at temperatures near
zero Kelvin and further studies, both experimental and theoretical,
will demonstrate the uniqueness of ultracold chemistry.
It may be expected that selection rules are more pronounced
and branching ratios of chemical reactions enhanced at ultralow
temperatures \cite{selection}. The study of \emph{ultracold chemistry} will take us into a
strange new world. Even the smallest activation energy will surely exceed
the  available thermal energy. However, the large de Broglie wavelength
pertaining to the ultracold regime entirely changes the nature of
reaction dynamics. At such low temperatures, even the collisions of large
molecules exhibit significant quantum effects. Energy barriers on the
potential energy surface play a different role because, in the
hyperquantum regime, tunneling becomes the dominant reaction pathway.
Since tunneling and resonances are characteristic of this regime, they can
serve as ultra-sensitive probes of particular features of the potential
energy surface.

Interactions of molecules at low temperatures can be manipulated by
external electromagnetic fields and the creation of cold and ultracold
molecules may lead to the development of the field of controlled
chemistry.  External-field control of chemical reactions can be based on
several principles. The Zeeman and Stark effects may remove some of the
energetically allowed reaction paths or they may open closed reaction
channels \cite{roman_prl}, leading to suppression or enhancement of the reaction
efficiency. External fields couple the states of different total angular
momenta, so that forbidden electronic transitions may become allowed in
an external field and the transition rate may be controlled by the field
strength \cite{roman_shaped}.  As the F + H$_2$ reaction shows, the rate of  low temperature
abstraction reactions may be dramatically enhanced by  the presence of a
resonance state near threshold. By shifting molecular energy levels with
external fields, it should be possible to bring an excited bound level of
the reactive complex in resonance with the collision energy \cite{bodo_c3}. 
External fields may influence statistical properties of ultracold molecular gases
such as diffusion.

Achieving \emph{quantum degeneracy} in an ultracold ensemble of polar
molecules will greatly expand the scope for study and applications of
collective quantum phenomena in this new form of matter. Of primary
interest are the effects of rotational and vibrational excitation and
their manifestation in collisional properties. In particular, a BEC of
polar molecules would represent a system of relatively strongly
interacting particles. This could be used in elucidating the link
between  BECs in gases and in liquids. The study of ultracold fermionic
molecules is also of great interest: the electric dipole-dipole
interaction, which is predicted to be energy independent, may give rise to
a molecular superfluid via BCS pairing. This may take place at phase space
parameters achievable by evaporative cooling. Ultracold molecules could be
also used in \emph{quantum computation}. The scheme invokes ultracold
ground-state molecules trapped in a one-dimensional optical lattice
combined with an inhomogeneous electrostatic field \cite{demille02}. 

There is also a class
of experiments with cold molecules that could answer questions reaching
beyond molecular physics and chemistry: these experiments test
fundamental symmetries in nature, such as the time-reversal symmetry and
the symmetrization postulate. A possible violation of time reversal
symmetry may be revealed by measuring the electric dipole moment (EDM) of
the electron in heavy diatomic molecules. It is particularly thrilling that the
current experimental upper bound on EDM is close to several
predictions made by theories that go beyond the Standard Model
\cite{doyle_hinds1,doyle_hinds2,commins,pbo,regan,hudson,review_of_hinds}.
 An increase in accuracy 
by even one order of magnitude would therefore have a
dramatic impact on theory. Also rapid progress in testing the Pauli
principle for bosons with cold molecules (with zero-spin nuclei) is
expected.
Another fundamental experiment tests parity violation, via high-resolution 
spectroscopy of enantiomers of chiral molecules \cite{chardonnet}. The 
accuracy of this experiment is expected to be significantly enhanced by 
using cold molecules.

{\bf Acknowledgments}

\noindent
We acknowledge the Institute for Theoretical Atomic, Molecular and Optical Physics 
(ITAMP) at Harvard University and the Smithsonian Astrophysical Observatory and 
the Center for Ultracold Atoms (CUA) at Harvard University and Massachusetts Institute of Technology 
for supporting the 2004 Workshop on ultracold polar molecules. 
We thank John Bohn for his help on the Feshbach resonance section
and Kate Kirby for useful comments on the manuscript.  
We thank Franco Gianturco for encouraging the idea to organize this issue and 
the European Physical Journal D for the opportunity to publish this sizable
collection of papers.

\end{document}